\documentstyle[prl,twocolumn,aps,epsf,rotate]{revtex}
\begin{document}
\draft

\title{Self-organized evolution in socio-economic environments}

\author{ A. Arenas$^*$, A. D\'{\i}az-Guilera$^{\dagger}$, C. J.
P\'{e}rez$^{\dagger}$ and F. Vega-Redondo$^{\ddagger}$ }

\address{\em $^*$ Departament d'Enginyeria Inform\`{a}tica,
Universitat Rovira i Virgili, Carretera Salou s/n, E-43006
Tarragona, Spain.\newline $^{\dagger}$ Departament de F\'{\i}sica
Fonamental, Universitat de Barcelona, Diagonal 647, E-08028
Barcelona, Spain.
\newline $^{\ddagger}$ Facultad de Econ\'{o}micas, Universidad de
Alicante, E-03071 Alicante, Spain. }

\date{\today}
\maketitle


\begin{abstract}
We propose a general scenario to analyze social and economic
changes in modern environments. We illustrate the ideas with a
model that incorporating the main trends is simple enough to
extract analytical results and, at the same time, sufficiently
complex to display a rich dynamic behavior. Our study shows that
there exists a macroscopic observable that is maximized in a
regime where the system is critical, in the sense that the
distribution of events follow power-laws. Computer simulations
show that, in addition, the system always self-organizes to
achieve the optimal performance in the stationary state.
\end{abstract}


\narrowtext

The evolution of socio-economic environments is attracting the interest
of the physics community due to the inherent complexity of many dynamic
processes. In particular, concepts and tools widely used in
nonequilibrium statistical physics have proved to be quite useful
when studying the complex behavior of interacting economic
agents \cite{santafe,bouchaud,econophys}.

There are clear evidences that social and economic change in
modern societies typically come in ``waves'' with seemingly little
intertemporal structure. There are many factors that can
contribute to such complex evolution but, in essence, any theory
able to account for the inherent dynamics of the phenomenon should
consider how the stimulus for change spreads by gradual local
interaction through a social network as well as the incentives
that govern individual behavior
\cite{blume,ellison,mailath,young}. The hope is that,
independently of the particular choice for the microscopic rules
describing the dynamic behavior of the agents that form an
arbitrary system, one should observe some collective trends that
could be reflected in terms of macroscopic observables.

In this Letter, our main goal is to define a general scenario that
could be useful to understand evolution in socio-economic
environments and within such a broad field our concern is related
to technological progress. In a general sense, let us consider a
population of agents each of them interacting with a group of
neighbors in order to carry out projects of mutual interest. From
these collaborations agents obtain payoffs which, of course, tend
individually to be as large as possible. To be more precise these
{\em payoffs} should reflect several basic properties. First, they
should account for a basic benefit obtained just for having a
certain technological level. It might be thought as an index for
the technological potential productivity. It is reasonable to
assume that the higher the technological level the larger the {\em
base payoff} will be. Furthermore, it should measure how similar
the tools required to undertake a mutual project are. It should
favor those collaborations where both technological levels are
very similar (high compatibility) and punish any waste of
resources derived from a possible mismatch between them. In other
words, technological compatibility should induce high values of
the {\em payoff function }while significant costs should arise
from any degree of incompatibility \cite{farrel}. It is also
reasonable to assume that those costs are bounded from below (the
bankrupt).

The dynamics must be consequent with the aforementioned basic
trends. Two main ingredients contribute to the dynamical
evolution. One is the interaction with the rest of the population.
Each agent should have the possibility to modify her technological
level if the benefits derived from this change are increased. With
only this term the system might reach a quiescent state
where all the agents are happy with their respective technological
level, not necessarily the same for all of them. To complete the
picture, it is also natural to think of individual mechanisms of
technological improvement which could be modeled as a sudden
update of the state of a given agent. This change plays the role
of a perturbation and admits several interpretations (e.g. local
innovation, a shock in payoffs, population renewal, etc.).
Immediately, her nearest neighbors check whether an update to a
new technological state is more profitable for them. The process
can be extended all over the network triggering a wave of change
or avalanche till a new quiescent state is reached. Then, the
sequence of events is repeated again. Notice that in modern
socio-economic environments, the diffusion of
information/technology is usually a fast process while advances
are developed in a much slower time scale. Therefore, it is
reasonable to assume that both processes are defined in different
time scales. Other ingredients can also be incorporated into this
general framework but, up to now, let us keep this simple picture
in mind.

Next point concerns the characterization of progress. More
precisely, we need to measure things in terms of a macroscopic
observable. Accounting for the rate of advance, the most natural
choice is the mean velocity of progress, which can be defined as
\begin{equation}
\rho =\frac{<H(t)>}{<s(t)>},  \label{rho}
\end{equation}
where $s$ is the number of agents involved in one of these
avalanches and $H$ is the total advance induced by the wave on the
whole population. Brackets denote average over time. Assuming that
each individual update induces a certain cost, $\rho$ also
measures the total cost needed to reach a certain global
(averaged) technological level. For instance, a high advance rate
will imply to achieve some given level at minimum total cost, i.e.
with the minimal number of individual updates.

Several questions arise in a natural manner. For a given system
defined in such scenario, is there any kind of behavior which
could optimize $\rho$?. If this is the case, how does the system
evolve towards the optimal state?.

Let us steer the intuition of the reader with a physical
discussion on the basis of the scenario proposed so far. If the
cost of an update is very small, then all agents are willing to
adopt the best available technology and any new particular
improvement will immediately be diffused through the whole
population, leading to an avalanche involving a large number of
individuals, eventually the whole system. In contrast, in the
opposite case it is generally difficult to find an agent
interested in changing her current state since even if a rather
advanced technology is available, the cost will typically be too
high leading to a situation where avalanches tend to be very
small. These two extreme situations can be identified with a
supercritical or subcritical regime, respectively. In both cases
the advance rate defined in (1) tends to be independent of the
number of agents $n$. From a theoretical standpoint one might
expect that in the intermediate range the most interesting
phenomena can emerge since it is where rich dynamic behavior may
flourish. In this regime, we expect that a substantial degree of
heterogeneity (but one that can be eventually broken by the advance of
technological avalanches) plays a fundamental role.

Once one knows the distribution of $s$ and $H$ then it is
straightforward to work out $\rho$ and determine if there is any
specific regime where the performance of a given system is
optimal. There are plenty of evidences reported in the literature
showing that quantities like $s$ and $H$ follow power law
distributions
\cite{santafe,bak,jensen,krugman,firms,rednerbis,redner,sornette}.
Therefore, we assume that the avalanche-size distribution obeys a
power law $P(s)\sim 1/s^{\gamma }$, for some $\gamma >0$, as well
as the distribution of technological advances per avalanche
$P(H)\sim 1/H^{\beta }$ for some positive $\beta$. In addition, we
take the natural assumption of considering that avalanche sizes
and induced advances should also be related, on average, through
some power relationship of the form:
\begin{equation}
H\sim s^{\alpha }  \label{power3}
\end{equation}
with $\alpha \geq 1$. Notice that the lower bond $\alpha=1$
corresponds to the two aforementioned extreme situations: either a
uniform growing front (supercritical) or hardly interacting agents
(subcritical). It can be easily shown that, provided $\gamma $ and
$\beta $ are larger than 1, the following relation should hold
among the exponents \cite{jensen}:
\begin{equation}
\alpha =\frac{\gamma -1}{\beta -1}.
\label{scalinglaw}
\end{equation}
It is also straightforward to find that the rate of
technological progress is
\begin{equation}
\rho =\frac{2-\gamma }{\alpha -\gamma +1}
\frac{n^{1+\alpha -\gamma }-1}{n^{2-\gamma }-1},
\label{rhodegamma}
\end{equation}
where $n$ is the size of the system (number of agents).
For large $n$, three different
regimes can be considered:

\noindent A) $\gamma <2$:
\[
\rho =\frac{2-\gamma }{\alpha -\gamma +1}n^{\alpha -1},
\]

\noindent B) $2<\gamma <\alpha +1$:
\[
\rho =\frac{\gamma -2}{\alpha -\gamma +1}n^{\alpha +1-\gamma},
\]

\noindent C) $\gamma >\alpha +1$:
\[
\rho =\frac{\gamma -2}{\gamma -\alpha -1}.
\]

Several conclusions can be extracted from these expressions.
First, notice that there are two regimes where the rate of
technological advance increases with the size of the system which
suggests that large economic ensembles enjoy beneficial "scale
effects", i.e. large economies grow faster. It is in these
regions where one could expect an optimal advance rate and
therefore an
optimal collective performance. In contrast, there is another
regime where $\rho$ is independent of $n$ suggesting a poorer
cooperative dynamics.

Let us illustrate the general scenario by proposing a particular
model endowed with the essential features requested in the
introduction. In this way we can investigate how the distribution
of technological advances depends on microscopic details of the
dynamics. In order to enhance transparency we have reduced the
complexity of the model as much as possible. We have considered a
system formed by a population of {\it n} agents defined on a periodic
1D geometry and nearest neighbor interactions. Each agent is
characterized by a positive real variable $a_{i}(t)$ identified as
the technological level. The interaction with the neighbors is
evaluated in terms of the payoff function that we have chosen to
be
\begin{equation}
\psi (a,a')= \left\{
\begin{array}{ll}
a-k_1(1-e^{-(a-a')})& \mbox{if } a>a'\\
a-k_2(1-e^{-(a'-a)})& \mbox{if } a<a'
\end{array}
\right.
.
\label{payoffs}
\end{equation}
Thus, the base payoff obtained from using a certain technology is
assumed equal to $a$ while the incompatibility costs resulting
from being too advanced or too backwards relative to neighbors are
parametrized, respectively, by positive factors $k_{1}$ and
$k_{2}$. However, as we will see later, the
overall properties of the system only depend on the difference $%
k=k_{1}-k_{2} $.

The dynamics of the model has two main components. At each time
step a randomly selected agent is chosen to update her technological
level from
$a_{i}(t-1)$ to $a_{i}(t)=a_{i}(t-1)+\tilde{\sigma}_{i}(t)$ where
$\tilde{\sigma}_{i}(t)$ is a i.i.d. random variable. The $j=i\pm 1$
agent
now has three options: either to maintain her level or to adopt
the level of one of her two neighbors. She is assumed to take
that action $a=\{ a_j,a_{j-1},a_{j+1}\}$,
which maximizes her total
payoff $\psi(a,a_{j-1})+\psi(a,a_{j+1})$ \cite{preparation}.
This process continues until no agent wants to perform any
adjustment in her technological level. Then, again one agent is
updated randomly and so on.

\begin{figure}
\epsfclipon
\epsfxsize=0.8\linewidth
\rotate[r]{\epsffile{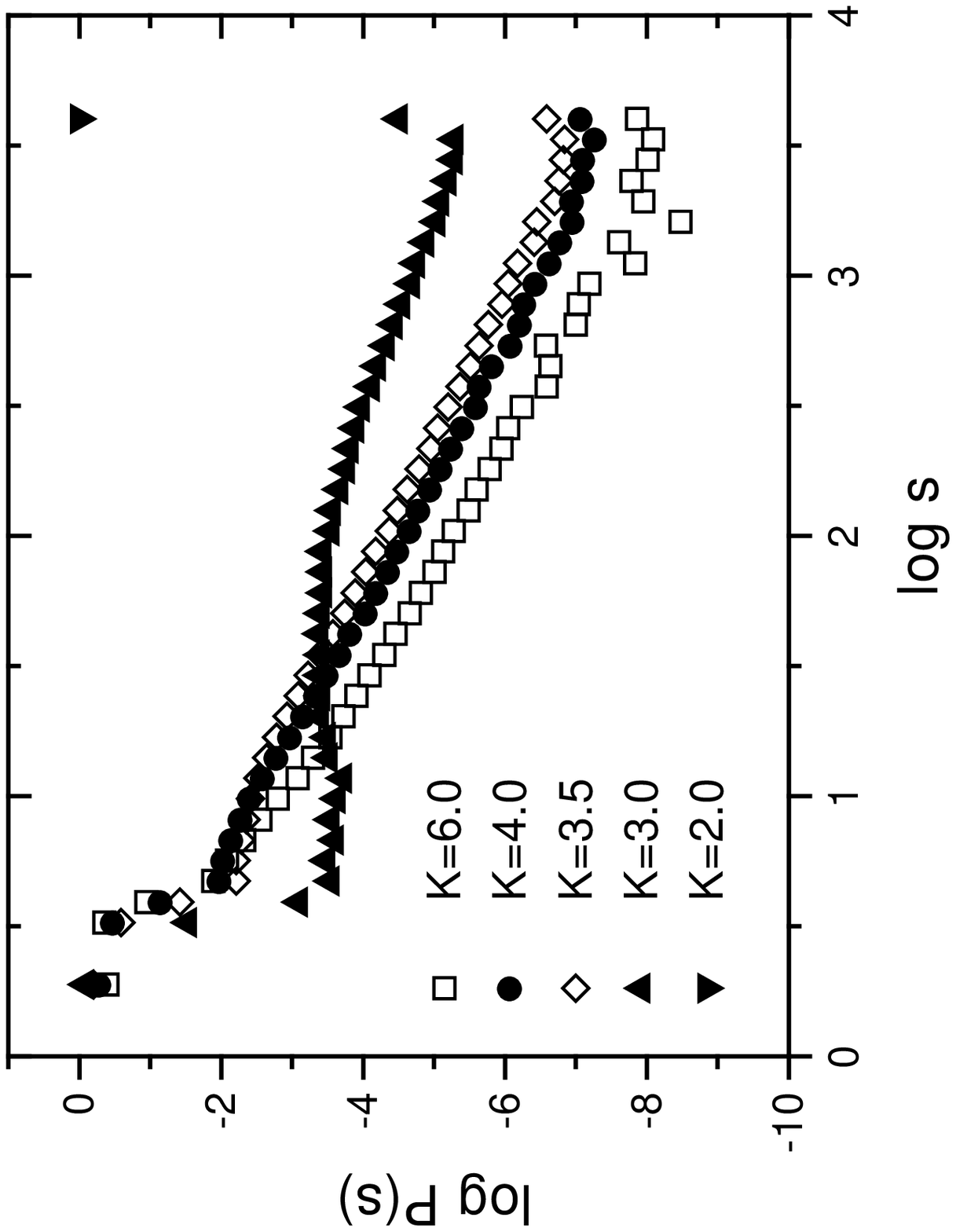}}
\caption{Logarithm of the probability of having an
avalanche of size $s$ vs. logarithm of the size $s$, for different
values of $k=k_{1}-k_{2}$. The length of the system is kept fixed
at $n=4096$. For $k=2$ only events of size $n$ are observed.
}
\end{figure}

Three different regimes are clearly observed in simulations. Fig.
1 shows the distribution of avalanches for different values of the
parameter $k$. As we expected, below a certain critical value all
the avalanches are of the size of the system. The technology
advances at unison like a uniform front. Any perturbation in the
system due to innovations is incorporated immediately by the rest
of agents because the cost of the update is very small. In the
opposite situation, i.e for large $k,$ the avalanches are
of small size. In the limit $k\rightarrow\infty$ the agents behave
independently and hence avalanches
are of size one; this
corresponds to the random deposition
model well known in surface growth \cite{barabasi}.
For intermediate values of
the coupling parameter the distribution of avalanches follows a
power law for several orders of magnitude of avalanche sizes.
Therefore, in this regime there is a clear absence of time as well
as length scales typical of a critical state \cite{bak,jensen}.

A deeper analysis of the model shows that it is possible to
extract analytical information about the location of the
supercritical regime. In particular, by only using local
arguments, it is
straightforward to show that if the difference in technological
level between two neighboring sites of a given agent $i$,
denoted by $\Delta=  \| a_{i+1}-a_{i-1} \|$, satisfies
\begin{equation}
k\equiv k_{1}-k_{2}<k^{\ast }(\Delta )\equiv \frac{2\Delta
}{1-e^{-\Delta }},
\end{equation}
then agent $i$ will always choose the highest local technological
level which in its turn can trigger additional updates in
neighboring sites \cite{preparation}.
In our case, where perturbations are
assumed uniformly random distributed in the interval [0,1] and
consequently the difference between agents is a continuous
variable, when $k\leq k^{\ast }(0)=2$ any local inhomogenenity
cannot be sustained and the system achieves a global synchronized
state where the technological level of all the agents is exactly the same.

\begin{figure}
\epsfclipon
\epsfxsize=0.9\linewidth\epsffile{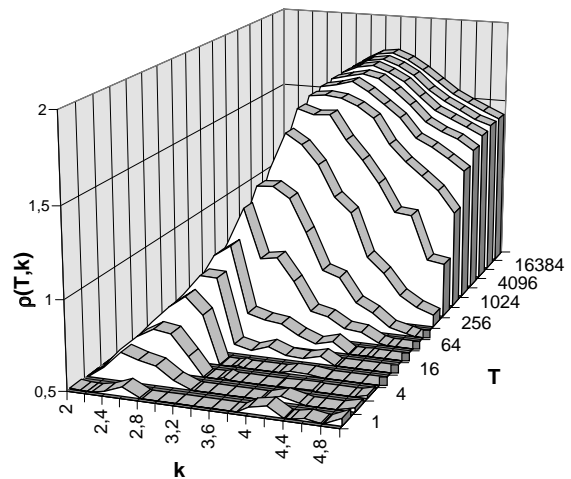}
\caption{Time evolution of $\protect\rho $ as a
function of $k$. The number of updates, $T$, accounts for the slow
time-scale. The system size is kept fixed at $n$=512. The result
is an average over 100 independent runs.}
\end{figure}

Now, let us consider the evolution of the macroscopic observable
$\rho$. Fig. 2 shows the time evolution of $\rho $ for different
values of $k$ and a fixed system size $n=512$, whereas Fig. 3
displays the stationary (long-run) values of $\rho $ for different
values of $k$ and different system sizes. Two important and
appealing features must be singled out here. First, Fig. 2 shows
that $\rho $ grows monotonically over time, the system
self-organizes to achieve, for each $k$ the best associated
performance. No matter what is the initial condition the system
evolves to maximize the advance rate. This quantity is maximal
when the stationary state is reached.  As far as we know, this is
the first model where critical behavior is attained through a
process of self-organization that maximizes a certain macroscopic
observable. Second, Fig. 3 indicates that $\rho $ is maximized
within the critical region, at a point located on its ``lower
edge'' \cite{kauffman} (i.e. within the narrow range where $k\in
\lbrack 3,4])$. Notice that in the two limit cases $k \leq 2$ and
$k \rightarrow \infty$, $H \approx s$ and therefore $\rho$ is
equal to the expected value of the external random perturbation,
$0.5$ in our case. Furthermore, these figures clearly show that
the advance rate
depends positively on the number of agents, $n$, stressing 
again the faster growth of large economies.
This fact is the motivation to investigate also if the model
presented in this
work accomplishes the relationship between exponents predicted
in the
general scenario. We have indeed confirmed that the
scaling relation (\ref{scalinglaw}) is fulfilled within
numerical accuracy, by fitting straight lines to the wide region
where power-laws are observed \cite{preparation}.

\begin{figure}
\epsfclipon
\epsfxsize=0.9\linewidth
\epsffile{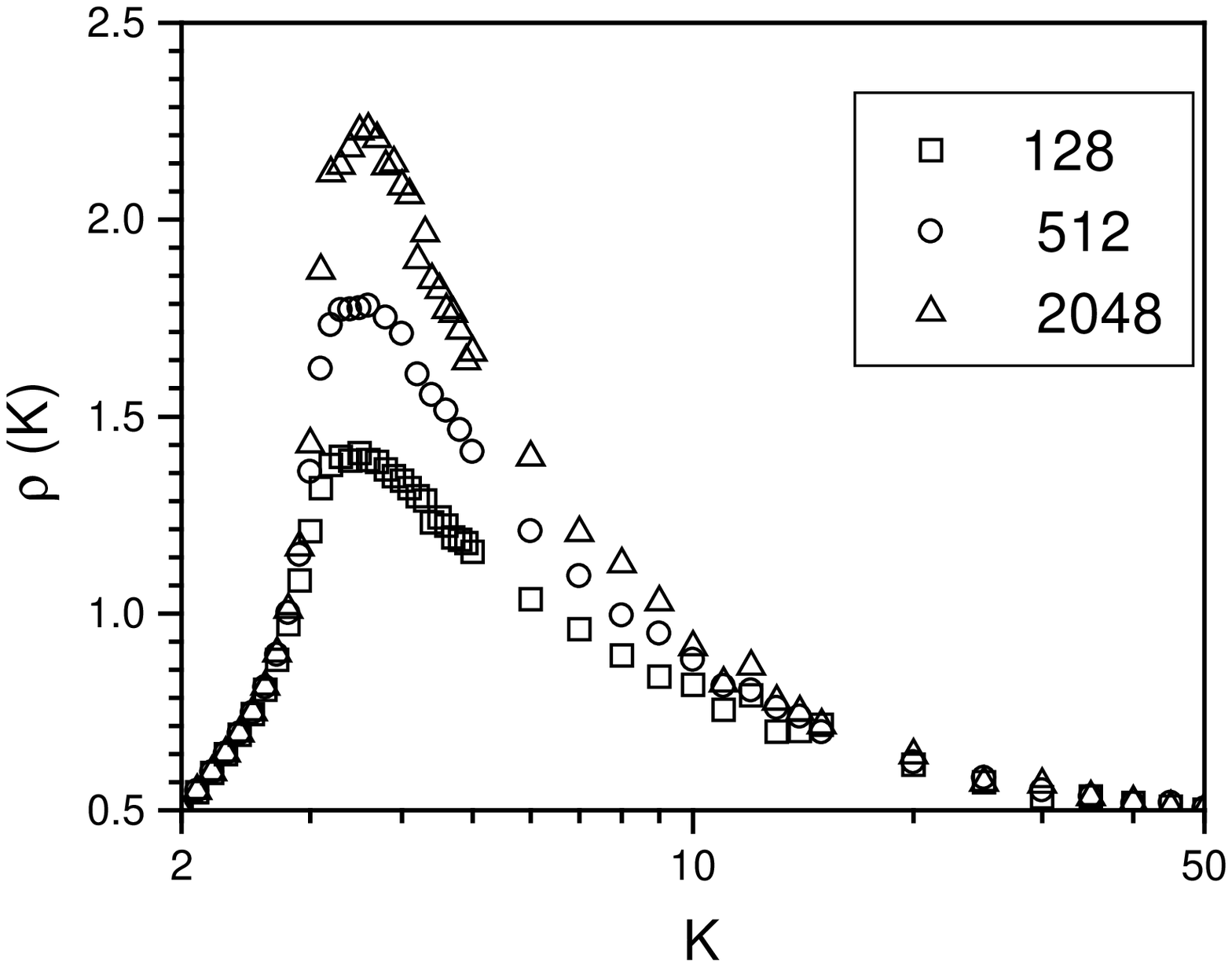}
\caption{Advance rate as a function of $k$, in a
log-log scale, for three different values of the length. For each
run we have generated $64\ast n$ avalanches and averaged over 10
independent realizations of the noise, except for $n=2048$ where
only two independent realizations have been considered.
} \label{rhodek}
\end{figure}

In conclusion, we have presented here a general scenario for the
study of evolution in socio-economic environments putting special
attention on the subject of technological progress. It is quite
appealing to realize that in a very general manner the framework
described in this Letter is able to predict the existence of
different regimes depending on the cost associated to the
improvement/diffussion of technology and that these regimes can be
computed directly from a macroscopic quantity without specifying
details about the underlying microscopic dynamics and payoff
functions. Even more, we have shown through a simple model that
critical behavior is attained in a natural way through a process
of self-organization that maximizes a macroscopic observable: the
advance rate.

\acknowledgments
The authors are gratefully acknowledged to A. Mas-Colell and H.J.
Jensen for helpful comments.
This work has been supported by DGES of the Spanish Government,
grants PB96-0168 and
PB97-0131, and EU TMR grant ERBFMRXCT980183.

\end{document}